\documentclass[conference]{ieeeconf}      

\IEEEoverridecommandlockouts                             
\usepackage{graphicx}      
\usepackage{amssymb,amsmath,mathrsfs}
\usepackage{comment}

\usepackage{flushend} 
\usepackage{balance} 

\newcommand{\R}{\mathbb{R}}

\usepackage{algorithm}
\usepackage{algpseudocode}
\algnewcommand\algorithmicinput{\textbf{input:}}
\algnewcommand\INPUT{\item[\algorithmicinput]}
\algnewcommand\algorithmicparam{\textbf{parameter:}}
\algnewcommand\PARAMETER{\item[\algorithmicparam]}
\algnewcommand\algorithmicoutput{\textbf{output:}}
\algnewcommand\OUTPUT{\item[\algorithmicoutput]}
\algnewcommand\FUNC{\item[\textbf{function}]}
\long\def\@makealgocaption#1#2{\vskip 2ex \small
  \hbox to \hsize{\parbox[t]{\hsize}{{\bfseries #1.} #2}}}
\usepackage[hoptionsi]{overpic}
\usepackage{bbm}

\usepackage{algorithmicx}
\usepackage{algorithm}
\usepackage{algpseudocode}

\usepackage{epsfig}
\usepackage{multirow}

\usepackage{amsthm}
\usepackage{amssymb}
\usepackage{amsmath}
\usepackage{bm}
\usepackage{cite}
\usepackage{breqn}
\usepackage{float}
\usepackage{mathrsfs}
\usepackage{mathtools}
\usepackage{amsfonts}
\usepackage{graphicx}
\usepackage{algorithm, algpseudocode}

\usepackage[usenames]{color}
\usepackage{xcolor}
\usepackage{amsfonts}
\usepackage{relsize}
\usepackage{lmodern}
\usepackage{slantsc}
\usepackage[mathscr]{eucal}
\usepackage{enumerate}
\usepackage[final]{pdfpages}
\usepackage{stfloats}
\usepackage{blindtext}

\usepackage{caption}
\usepackage{balance}
\usepackage{flushend}
\usepackage{epsfig}

\usepackage{multirow}

\usepackage{amsthm}
\usepackage{amssymb}
\usepackage{amsmath}
\usepackage{bm}

\usepackage{cite}

\usepackage{breqn}
\usepackage{float}
\usepackage{mathrsfs}
\usepackage{mathtools}
\usepackage{amsfonts}
\usepackage{graphicx}

\usepackage{algorithm, algpseudocode}

\usepackage[usenames]{color}
\usepackage{xcolor}
\usepackage{amsfonts}
\usepackage{relsize}
\usepackage{lmodern}
\usepackage{slantsc}
\usepackage[mathscr]{eucal}

\usepackage{dsfont}
\usepackage{bbm}
\usepackage{bm}

\usepackage{caption}
\usepackage{tabularx, booktabs}
\usepackage{adjustbox}

\usepackage{xcolor}

\usepackage{colortbl}
\usepackage{lipsum}
\usepackage{scalerel}
\usepackage{stackengine}
\usepackage{tabularx,ragged2e}
\usepackage{booktabs}

\usepackage[utf8]{inputenc}
\usepackage{array}
\usepackage{makecell}
\let\labelindent\relax
\usepackage{enumitem}
\usepackage{makecell}
\usepackage{multicol}
\usepackage{scalerel}
\usepackage{subcaption}

\usepackage[font={small}]{caption}

\def\equ{\thinspace{=}\thinspace}

\usepackage{accents}

\DeclareMathAlphabet{\mathpzc}{OT1}{pzc}{m}{it}
\DeclarePairedDelimiterX{\norm}[1]{\lVert}{\rVert}{#1}

\algnewcommand\Input{\item[\hspace{6pt}\textbf{Input:}]}
\algnewcommand\Output{\item[\hspace{6pt}\textbf{Output:}]}
\algnewcommand\OutputVal{\textbf{output} }
\newcolumntype{L}[1]{>{\raggedright\arraybackslash}p{#1}}
\newcolumntype{C}[1]{>{\centering\arraybackslash}p{#1}}
\newcolumntype{R}[1]{>{\raggedleft\arraybackslash}p{#1}}

\title{\bf  Koopman-Based Dynamic Environment Prediction for Safe UAV Navigation}

\author{ Vitor Bueno, Ali Azarbahram,  Marcello Farina, and  Lorenzo Fagiano 
  \thanks{This work has been supported by the Italian Ministry of Enterprises and Made in Italy in the framework of the project 4DDS (4D Drone Swarms) under grant no. F/310097/01-04/X56.}
\thanks{V. Bueno, M. Farina and L. Fagiano are with the Dipartimento di Elettronica, Informazione
 e Bioingegneria, Politecnico di Milano, 20133 Milan, Italy. (e-mail: \{vitor.jardim, marcello.farina, and lorenzo.fagiano\}@polimi.it).
 A. Azarbahram is with The Department of Electrical Engineering, Chalmers University of
Technology, Gothenburg, 412 96, Sweden. (e-mail: ali.azarbahram@chalmers.se). 
  }
  \thanks{ $\ast$ V. Bueno and  A. Azarbahram contributed equally to this work.}
   }

\begin{document}

\maketitle
\thispagestyle{empty} 
\pagestyle{empty} 

\begin{abstract}
This paper presents a Koopman-based model predictive control (MPC) framework for safe UAV navigation in dynamic environments using real-time LiDAR data. By leveraging the Koopman operator to linearly approximate the dynamics of surrounding objets, we enable efficient and accurate prediction of the position of moving obstacles. Embedding this into an MPC formulation ensures robust, collision-free trajectory planning suitable for real-time execution. The method is validated through simulation and ROS2-Gazebo implementation, demonstrating reliable performance under sensor noise, actuation delays, and environmental uncertainty. 
\end{abstract}



\section{Introduction}
\label{Sec. 1}

Autonomous navigation for unmanned aerial vehicles (UAVs) is a fundamental capability that enables drones to operate independently in complex, dynamic settings without continuous human oversight \cite{bagloee2016autonomous, saccani2022multitrajectory}. Achieving autonomy requires addressing multiple challenges, such as generating trajectories that respect actuator limitations, avoid collisions, and are computationally feasible for real-time deployment \cite{tang2018autonomous}.
When the environment is fully known in advance, including obstacle locations and geometry, well-established path-planning strategies exist \cite{tang2018autonomous, chen2016uav, chengjun2017spare}, often incorporating Model Predictive Control (MPC) \cite{bemporad2009hierarchical, mayne2005robust, bemporad2011decentralized}, which is particularly suited to such problems due to its predictive structure and natural incorporation of constraints.

However, many real-world scenarios involve partially known or dynamically changing environments. In these cases, UAVs must rely on onboard sensors to interpret their surroundings and adapt their motion accordingly. This necessitates real-time obstacle detection and avoidance, balancing safety with efficiency in reaching a target. To address this, many methods use exteroceptive sensor data to either reconstruct a local map \cite{liu2017planning} or perform immediate obstacle avoidance directly from raw measurements \cite{park2020collision}.
Among the most critical challenges in these environments is the presence of moving obstacles, such as other vehicles, people, or objects, which require predictive capabilities to ensure collision-free navigation. Enabling UAVs to detect, anticipate, and avoid such dynamic threats is essential for safe and autonomous operation, especially in urban areas, disaster zones, or densely populated settings.

Recent research has explored MPC-based solutions for dynamic obstacle avoidance. For instance, a chance-constrained MPC enhanced with control barrier functions is introduced in \cite{li2023moving}, while \cite{olcay2024dynamic} employs Gaussian Process regression for predictive modeling of obstacle motion. A risk-aware MPC with bootstrapped trajectory forecasts is proposed in \cite{wei2022moving}. The authors in \cite{lindqvist2020nonlinear} design a Nonlinear MPC approach incorporating real-time obstacle prediction and constraint management. A comprehensive review is presented in \cite{wang2024review}, which subdivides dynamic obstacle avoidance into three core components: environmental perception, trajectory prediction, and trajectory planning.

To address the challenge of modeling and controlling nonlinear systems such as UAVs and their environment, Koopman operator-based approaches have gained attention as a powerful data-driven framework. 
%
%
%
%
Several recent works have demonstrated the effectiveness of Koopman-based approaches in control. The method in \cite{gutow2020koopman} applies the Koopman operator to enforce chance constraints under uncertainty for efficient motion planning. In \cite{comas2021self}, the operator is used to model moving object trajectories from video sequences. A robust MPC design integrating environment-switching via Koopman models is proposed in \cite{oh2024koopman}. Robust tube-based MPC relying on finite-dimensional Koopman approximations is studied in \cite{zhang2022robust}, while \cite{wang2022data} combines neural networks and Koopman modeling within a constraint-tightened MPC. Finally, \cite{manzoor2023vehicular} provides an extensive overview of Koopman operator theory applied to system identification and control in mobility and automotive systems.

The Koopman operator is particularly beneficial for predicting the motion of dynamic obstacles, as it enables linear forecasting from high-dimensional data, thus providing a structured and tractable means of handling nonlinear and uncertain behaviors. This capability is leveraged in \cite{lu2024vector}, where offline Koopman learning is used to model obstacle dynamics for improved planning.
These predictive capabilities align well with MPC, enhancing the controller’s ability to foresee and avoid moving obstacles by planning safe and efficient paths. 
Consequently, integrating Koopman-based models into the MPC framework can significantly improve the autonomy, safety, and reliability of UAV systems in dynamic environments.
Motivated by these insights, we propose a data-driven approach that integrates Koopman-based prediction of moving obstacles into a convex MPC framework that marks to the best of our knowledge the first use of Koopman operators for dynamic environment modeling in autonomous UAV navigation.
The obstacle avoidance constraints are formulated as linear inequalities, ensuring convexity and enabling efficient optimization. The complete system is validated in a ROS2-Gazebo simulation, showcasing its practical feasibility, real-time performance, and potential for deployment in realistic UAV applications.

The remainder of this article is structured as follows: Section~II details the UAV model used; Section~III introduces the Koopman-based prediction of moving obstacles; Section~IV details the proposed controller for obstacle avoidance; simulation studies are presented in Section~V; and conclusions with future directions are discussed in Section~VI.





  
\section{The UAV Model}
\label{Sec. 2}

The overall framework considered in this work consists of three main components: an autonomous UAV, a LiDAR-based perception system, and a dynamic environment populated by moving obstacles. The UAV operates under a hierarchical control architecture, where a high-level motion planner computes the velocity reference ${\bm{V_{ref}}(t)} = [V_{ref,x},  V_{ref,y},  V_{ref,z}]^{\rm T} \in \R^{3}$, which is subsequently tracked by a low-level velocity controller. For brevity, we omit details on this low-level controller, which follows the standard structure used in \cite{fagiano2017systems}. Any implementation that ensures stable flight and accurate tracking of reference commands is suitable. This controller typically runs at high frequency (between 50 and 100 Hz), enabling precise and responsive behavior.

The UAV dynamics, when considered together with the internal velocity feedback loop, can be well-approximated by linear dynamics. This allows us to formulate the UAV motion planning task using a linear time-invariant (LTI) model, facilitating the use of efficient quadratic programming (QP) solvers within our predictive control framework. We model the drone’s translational dynamics using a continuous-time second-order system with position and velocity feedback, described as follows:

\begin{flalign}\label{UAVDynamicsContinuous}
\kern-0.3em \left[\begin{matrix} 
\dot{\bm{p}}(t)  \\
\dot{\bm{v}}(t)   
\end{matrix} \right]  = 
\kern-0.3em \left[\begin{matrix} 
\bm{0}_{3\times3}    &    \bm{I}_3\\
\bm{0}_{3\times3}    &    -\bm{K_{vel}}
\end{matrix} \right] \kern-0.3em \left[\begin{matrix} 
{\bm{p}}(t)  \\
{\bm{v}}(t)   
\end{matrix} \right]   +     \kern-0.3em \left[\begin{matrix} 
\bm{0}   \\
\bm{K_{vel}} 
\end{matrix} \right]  \bm{V_{ref}}(t).
\end{flalign}

Here, ${\bm{p}}(t)$ and ${\bm{v}}(t)$ denote the 3D position and velocity vectors of the UAV in Cartesian coordinates. The matrix $\bm{K_{vel}} \in \R^{3\times3}$ represents the velocity control gain. Through system identification, this gain can be tuned such that the model’s response closely resembles that of the actual nonlinear UAV dynamics.
We discretize the continuous model in (\ref{UAVDynamicsContinuous}) with sampling period $T_s$, yielding the control-oriented LTI discrete-time model:

\begin{flalign}\label{UAVDynamicsDiscr}
\bm{x}(k+1) = A \bm{x}(k) + B \bm{u}(k).
\end{flalign}

\noindent
In this model, the state vector ${\bm{x}}(k) = [p_x(k),  p_y(k),  p_z(k),  v_x(k),  v_y(k),  v_z(k)]^{\rm T}$ encapsulates both position and velocity states, while the control input vector is ${\bm{u}}(k) = [V_{ref,x}(k),  V_{ref,y}(k),  V_{ref,z}(k)]^{\rm T} \in \R^{3}$. This formulation directly supports velocity references as inputs and allows the MPC controller to enforce constraints on position, velocity, and even acceleration.  For readers interested in the system identification process and parameter tuning of such models, we refer to \cite{saccani2022multitrajectory}.
\section{Koopman Operator-based Prediction of the Position of Moving objects}
\label{Sec. 3}
To enable accurate prediction of obstacle movements, we leverage the Koopman operator framework, a powerful data-driven method that lifts nonlinear system dynamics into a higher-dimensional space where their evolution becomes approximately linear. Rather than modeling the nonlinear state transitions directly, this approach applies a transformation which encode the system’s dynamics in the lifted space.
Assume the evolution of a nonlinear system is governed by $\vartheta(k+1) = f(\vartheta(k))$, where $\vartheta(k) \in \R^{n_{\vartheta}}$ is the system state. This approach is based on the definition of a set of so-called observables, i.e., a set of possibly nonlinear functions of the state $\vartheta$, i.e., $\psi(\vartheta) \in \R^{n_{\psi}}$, such that there exists a linear mapping $\mathcal{K}$ (the so-called Koopman operator) with the following property:


\begin{flalign} \label{KOOPMAN}
\psi(\vartheta(k+1)) = \mathcal{K} \psi(\vartheta(k)).
\end{flalign}

\noindent
Here, $\mathcal{K} \in \R^{n_{\psi} \times n_{\psi}}$, which captures the evolution of observables, can be estimated from trajectory data using as in \cite{mezic2005spectral, williams2015data, brunton2016discovering}, offering a linear model that reflects the underlying nonlinear dynamics.
We simulate realistic sensing with a 3D LiDAR attached to the UAV for detecting the obstacles in the environment. Each detection is stored in a buffer, creating a time history of observations for every obstacle. This history maintains a fixed length by discarding the oldest entries once the buffer is full. Once a sufficient amount of data of an obstacle is collected, its Koopman operator $\mathcal{K}$ is estimated \cite{williams2015data}, providing a learned model to predict future obstacle positions.
Using the estimated Koopman operator, we recursively compute the obstacle’s future positions over a defined prediction horizon. Starting from the most recent observation, the lifted state is propagated through the operator $\mathcal{K}$, yielding a full predicted trajectory. 

\section{The MPC Problem}
\label{Sec. 4}

This section presents the formulation of the proposed MPC problem designed for real-time UAV navigation with moving obstacle avoidance using Koopman-based prediction. The goal is to compute an optimal control sequence that ensures the UAV follows a desired trajectory while avoiding collisions with dynamic obstacles. To track a reference point $\bm{r} = [r_x,  r_y,  r_z]^{\rm T}$, we define the following stage cost:

\begin{flalign} \label{stagecost}
\gamma( \bm{x}_{\mu}(k),  \bm{r} ) =  \norm{  \bm{x}^{1:3}_{\mu}(k) - \bm{r} }_Q^2.
\end{flalign}

\noindent
Here, $\norm{ \bm{x} }_Q^2 = \bm{x}^{\rm T} Q \bm{x}$, where $Q \in \R^{3\times3}$ is a symmetric positive-definite weighting matrix. The subscript $\mu$ in $\bm{x}^{1:3}_{\mu}(k)$ refers to the $\mu$-steps ahead prediction of the first three components of vector $\bm{x}$ over the finite horizon $H$, with $\mu \in \mathbb{N}_{[0,H]}$. To derive the proposed MPC approach, we formulate a rather standard reference-tracking finite-time optimal control problem over this horizon as:

\begin{subequations}
\begin{flalign}
\label{MPC_1}
& J_L^{\ast}(\bm{x}(k), \bm{r})  =   \min_{\bm{u}_{0:H-1}(k)} \sum_{\mu=0}^{H} \gamma( \bm{x}_{\mu}(k),  \bm{r} ) ,  \\
& \text{s.t.} ~~   \nonumber \\
\label{MPC_2}
& \bm{x}_{\mu+1}(k) = A \bm{x}_{\mu}(k) + B \bm{u}_{\mu}(k) , \forall \mu \in \mathbb{N}_{[0,H-1]}, \\
\label{MPC_3}
& \bm{x}_{0}(k) =  \bm{x}(k),  \\ 
\label{MPC_6}
& - \Bar{\bm{v}}  \leq  \bm{x}^{4:6}_{\mu}(k) \leq  \Bar{\bm{v}} , \forall \mu \in \mathbb{N}_{[0,H]},  \\
\label{MPC_5}
& - \Bar{\bm{u}} \leq \bm{K_{vel}} (\bm{u}_{\mu}(k) - \bm{x}^{4:6}_{\mu}(k)) \leq \Bar{\bm{u}},  \forall \mu \in \mathbb{N}_{[0,H-1]},\\
\label{MPC_4}
& (\eta_{\mu, { j_{\mu,\max}^{[i]}  }}^{[i]})^{\rm T} (\bm{x}^{1:3}_{\mu}(k) - \bm{p}^{[i]}_{\mu}(k)) \geq d^{[i]}  , \forall \mu \in \mathbb{N}_{[0,H]}.
\end{flalign}
\end{subequations}


The constraints in \eqref{MPC_6} and \eqref{MPC_5} enforce limits on the velocity and acceleration, using the bounds $\Bar{\bm{v}}$ and $\Bar{\bm{u}}$, respectively.
We now explain the moving obstacle avoidance constraints in \eqref{MPC_4}. Let $M$ denote the number of dynamic obstacles in the environment, indexed by $i = 1, \ldots, M$. Given the $i$-th obstacle’s approximate position $\bm{p}^{[i]}(k)$ sensed via LiDAR at time $k$, we define the evolution of the Koopman-lifted state in the prediction horizon as:

\begin{flalign} \label{ObstacleDynamics}
\psi_{\mu+1}(\bm{p}^{[i]}(k)) &\equ \mathcal{K} \psi_{\mu}(\bm{p}^{[i]}(k)),  \forall \mu \in \mathbb{N}_{[0,H-1]}, \nonumber \\
\psi_{0}(\bm{p}^{[i]}(k)) &= \psi(\bm{p}^{[i]}(k)).
\end{flalign}

\noindent
From this lifted sequence $\psi_{\mu}(\bm{p}^{[i]}(k))$, we extract the predicted position of obstacle $i$ at step $\mu$ as $\bm{p}^{[i]}_{\mu}(k)$ for all $\mu$.
To facilitate collision avoidance, we represent each predicted obstacle position $\bm{p}^{[i]}_{\mu}(k)$ as a spherical region of radius $R^{[i]}$. The minimum distance constraint between the UAV and the $i$-th obstacle is defined as:


\begin{flalign} \label{DistanceTo}
\| \bm{x}^{1:3}_{\mu}(k) - \bm{p}^{[i]}_{\mu}(k) \| ~\geq~d^{[i]}.
\end{flalign}
\begin{flalign} \label{DistanceTo}
 d^{[i]}= R + R^{[i]} + \delta^{[i]}.
\end{flalign}

\noindent
Here, $R$ is the UAV's safety radius and $\delta^{[i]}$ is an extra margin ensuring a safe separation. However, this constraint is nonlinear and non-convex, making it unsuitable for direct use in a convex optimization problem.
To address this, we approximate the obstacle’s safety region using a regular polytope $\zeta_{\mu}^{[i]}$ characterized by $\phi_{\mu}^{[i]}$ half-space constraints. Each constraint represents a supporting plane of the polytope


\begin{flalign} \label{polytope}
(\eta_{j}^{[i]})^{\rm T} (\bm{x}^{1:3}_{\mu}(k) - \bm{p}^{[i]}_{\mu}(k)) \leq R^{[i]}, \quad j = 1, \dots, \phi_{\mu}^{[i]}.
\end{flalign}

\noindent
In this expression, $\eta_{\mu,j}^{[i]}$ is the outward-pointing normal vector of the $j$-th plane, and $d_{\mu}^{[i]}$ denotes its distance from the obstacle’s center. These constraints define a polytope that fully contains the obstacle’s spherical region, providing a linear approximation suitable for MPC.
To determine the most critical constraint for safety, we define the signed distance from the UAV’s position to each plane


\begin{flalign} \label{polytope2}
\rho_{\mu,j}^{[i]}  \equ  (\eta_{j}^{[i]})^{\rm T} (\bm{x}^{1:3}_{0}(k) - \bm{p}^{[i]}_{\mu}(k)) - R^{[i]}.
\end{flalign}

\noindent
We then select the plane with highest signed distance to the UAV

\begin{flalign} \label{polytopeMAX}
j_{\mu,\max}^{[i]} = \arg \max_{j \in \{1, \dots, \phi_{\mu}^{[i]}\}}  \rho_{\mu,j}^{[i]}.
\end{flalign}

\noindent
This corresponds to the hyperplane that best approximates the one tangent to the obstacle's sphere and perpendicular to the vector connecting the UAV at time $k$ to the obstacle at time step $\mu$. The linear constraint included in the MPC formulation to achieve safe navigation is then given by:

\begin{flalign} \label{LINEARCONSTRAINT}
(\eta_{\mu, { j_{\mu,\max}^{[i]}  }}^{[i]})^{\rm T} (\bm{x}^{1:3}_{\mu}(k) - \bm{p}^{[i]}_{\mu}(k)) \geq d^{[i]}.
\end{flalign}

\begin{figure}[t]
	\centering
	\includegraphics[scale = 0.7]{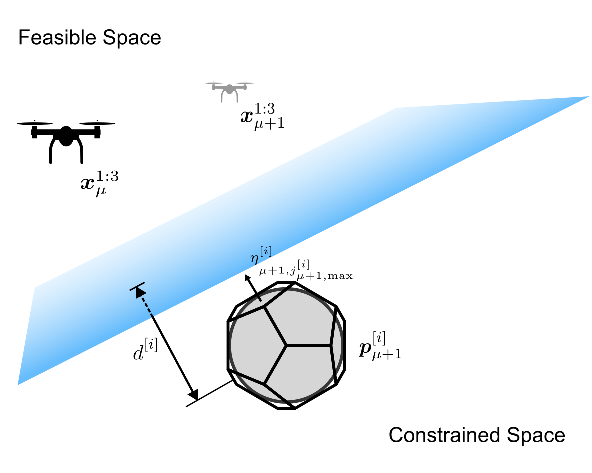}
	\captionsetup{font=footnotesize}
	\caption[]{Polygonal obstacle approximation and minimum distance constraint visualization. 	\label{FIG_dist_constraint}}
\end{figure}

\noindent
This constraint, inserted into \eqref{MPC_6}, ensures that the UAV remains on the safe side of the closest polytope plane, thereby preventing collisions with the $i$-th obstacle. This technique follows the approach introduced in \cite{PERIZZATO2015260}. Figure \ref{FIG_dist_constraint} illustrates the resulting feasible region for UAV motion based on predicted obstacle positions and the applied safety margins.

\noindent

\begin{figure}[b]
	\centering
	\includegraphics[scale = 0.3]{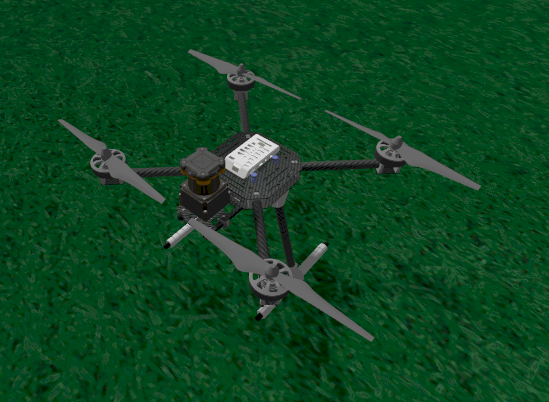}
	\captionsetup{font=footnotesize}
	\caption[]{Simulated quadcopter.  	\label{FIG_GZ}}
\end{figure}

\begin{figure}[t]
	\centering
	\includegraphics[scale = 0.9]{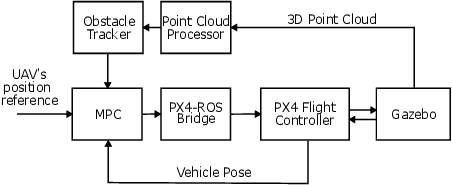}
	\captionsetup{font=footnotesize}
	\caption[]{Simplified interaction diagram of the ROS2 nodes relative to the main vehicle control. \label{FIG_nodes_diagram}}
\end{figure}


\begin{figure}[t]
	\centering
	\includegraphics[scale = 0.75]{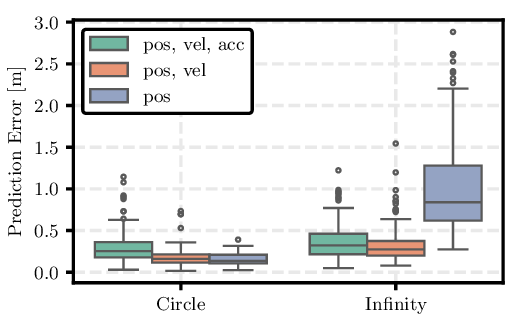}
	\captionsetup{font=footnotesize}
	\caption[]{Prediction performance on circular and infinity-shaped trajectories.
    \label{FIG_pred_results_1}}
\end{figure}

\begin{figure}[b]
	\centering
	\includegraphics[scale = 0.8]{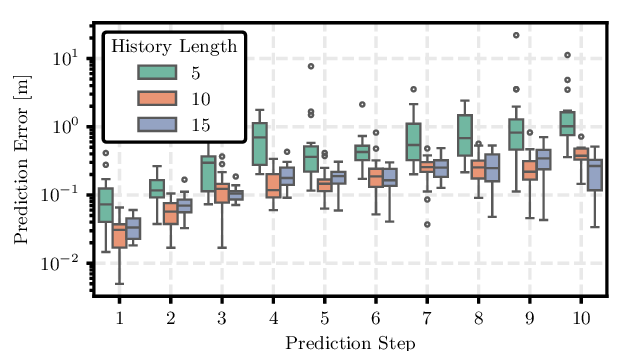}
	\captionsetup{font=footnotesize}
	\caption[]{Effect of prediction step and history length on prediction accuracy.
    \label{FIG_pred_results_2}}
\end{figure}

\begin{figure}[t]
	\centering
	\includegraphics[scale = 0.99]{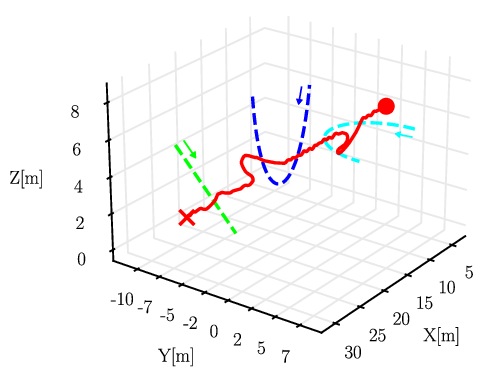}
	\captionsetup{font=footnotesize}
	\caption[]{Trajectory of the UAV starting from its initial position (solid red circle) towards the reference position (red cross) in the presence of 3 moving obstacles, whose trajectories are represented by the cyan, blue and green dashed lines. \label{FIG_3D_avoidance}}
\end{figure}

\begin{figure}[b]
	\centering
	\includegraphics[scale = 0.9]{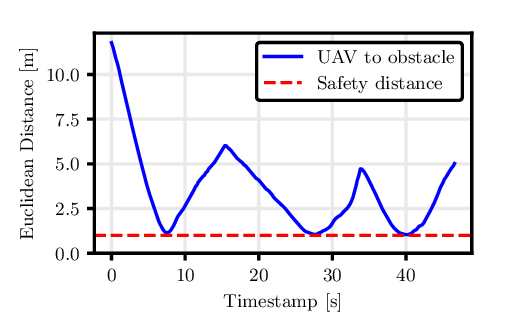}
	\captionsetup{font=footnotesize}
	\caption[]{Distance from the UAV to the nearest obstacle along the mission, together with the minimum distance imposed in the MPC formulation. \label{FIG_distance_to_obstacle}}
\end{figure}

\begin{figure}[b]
	\centering
	\includegraphics[scale = 0.75]{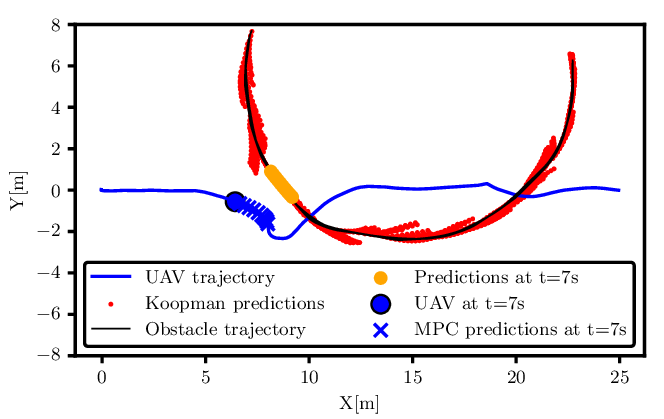}
	\captionsetup{font=footnotesize}
	\caption[]{Detailed avoidance and prediction performance on circular obstacle trajectory.
    \label{FIG_2D_avoidance}}
\end{figure}

\begin{table*}[t]
\centering
\caption{Prediction error comparison between the GP method and the proposed Koopman-based approach.}
\begin{tabular}{lcccccc}
\toprule
\multirow{2}{*}{Motion Type} & \multicolumn{3}{c}{GP \cite{olcay2024dynamic}} & \multicolumn{3}{c}{Koopman (Proposed)} \\
\cmidrule(lr){2-4} \cmidrule(lr){5-7}
& RMSE & MAE & MaxErr & RMSE & MAE & MaxErr \\
\midrule
Circular (XY plane) & 0.23& 0.20& 0.38& 0.17& 0.15& 0.27\\
Planar Figure-Eight Trajectory (XY plane)  & 0.55& 0.47& 0.99& 0.47& 0.38& 0.93\\
\bottomrule
\end{tabular}
\label{tab:error_comparison}
\end{table*}

\begin{table}[b]
\centering
\caption{Average computation time comparison.}
\begin{tabular}{@{}lr@{}}
\toprule
                 & Execution Time [ms] \\ \midrule
Koopman (Proposed)   & 0.42          \\
GP \cite{olcay2024dynamic} & 23.33         \\ \bottomrule
\end{tabular}
\label{tab:time_diff}
\end{table}

\section{Simulation results}
	\label{Sec. 5}

To validate the efficacy of the proposed Koopman-based MPC framework, we leverage a high-fidelity simulation environment that integrates Gazebo, ROS2, and the PX4 Firmware. These tools provide realistic UAV dynamics, support modular system design, and ensure accurate simulation of onboard control behavior. The setup enables comprehensive testing of the autonomous navigation system, including its capability to predict and avoid dynamic obstacles using a data-driven Koopman operator. All simulations were conducted on a computer with an AMD Ryzen 9 3900X 12-Core Processor @ 3.8 GHz.

Gazebo Garden is selected for its advanced 3D rendering and physics simulation capabilities, particularly useful for capturing UAV dynamics and obstacle interactions. ROS2 Humble acts as the middleware, offering robust communication infrastructure and real-time control capabilities. PX4-SITL (v1.14.3) complements the setup with an embedded flight controller, using a state estimator and inner-loop controller to convert velocity references into actuator-level commands. The default PX4 parameters were used to maintain standardization.
The simulation environment closely mimics realistic flight conditions. A flat ground plane at $z = 0$ acts as the only static obstacle. A standard adiabatic atmosphere model and gravity of $9.8 \, \mathrm{m/s^2}$ are used. The UAV, modeled as an x500 V2 quadrotor (Figure \ref{FIG_GZ}), has a mass of $2.0 \, \mathrm{kg}$ and a thrust capacity of $31.16 \, \mathrm{N}$. Sensor simulation includes a 3D LiDAR with a $20.0 \, \mathrm{m}$ range and Gaussian noise ($\mu = 0.0 \, \mathrm{m}$, $\sigma = 0.01 \, \mathrm{m}$), along with an IMU, GPS, and barometric sensor.

The control architecture is modularized into ROS2 nodes that communicate via topics at a rate of $T_s=0.2 \mathrm{s}$. A block-level diagram of the major nodes and communication flows involved in vehicle navigation is shown in Figure \ref{FIG_nodes_diagram}.
Referring to this figure, the "px4\_ros\_bridge" serves as the communication bridge between the PX4 flight controller and high-level ROS2 nodes. It handles UAV operations such as arming, takeoff, and mission execution by forwarding velocity commands.
The “perception\_module” processes incoming LiDAR data $\mathcal{S}(k)$ using the PCL (v1.12.1) library. It transforms the points in $\mathcal{S}(k)$ to the global coordinate system, filters out self-points and ground points using RANSAC \cite{fischler1981random}, then clusters the remaining data into $C(k)$ groups $\mathcal{S}^{[c]}(k)$, where $c =1, \dots, C(k)$ represents a detected obstacle.

The centroid and bounding sphere radius for each cluster are estimated using the following equations:

\begin{align}
     \bm{p}^{[c]}(k) &= \frac{1}{|\mathcal{S}^{[c]}(k)|} \sum_{\bm{p}^s \in \mathcal{S}^{[c]}(k)}{\bm{p}^s},
    \label{EQ_cluster_c} \\
    R^{[c]}(k) &= \max_{\bm{p}^s \in \mathcal{S}^{[c]}(k)}{\|\bm{p}^{[c]}(k) - \bm{p}^s\|},
    \label{EQ_cluster_r}
\end{align}



\noindent
This approach guarantees that each obstacle cluster is enclosed within a bounding sphere, ensuring a conservative yet efficient collision representation.
Still referring to  Figure \ref{FIG_nodes_diagram}, the “obstacle\_tracker” node associates each cluster's center and radius to a unique obstacle $\hat{\bm{p}}^{[i]}(k)$ and $\hat{R}^{[i]}(k)$ and maintains an obstacle history buffer as:

\begin{align}
    \mathcal{G}(k) &= \{\mathcal{H}^{[i]}(k)\}_{i=1}^{M(k)},
    \label{EQ_vect_hist} \\
    \mathcal{H}^{[i]}(k) &=\{\vartheta^{[i]}(\gamma)\}_{\gamma=k-O^{[i]}(k)T_{\kappa}/T_s}^k,
    \label{EQ_vect_hist} \\
     \vartheta^{[i]}(k) &=[\hat{\bm{p}}^{[i]}(k), \hat{\bm{p}}^{[i]}(k - T_{\vartheta}/T_s), \nonumber \\
     &~~~~~~~~~~~~~~~~~\hat{\bm{p}}^{[i]}(k - 2T_{\vartheta}/T_s)]^T,
    \label{EQ_vect_state}
\end{align}




\noindent
$T_{\vartheta}=3T_s$ is used and a Savitzky-Golay filter is applied to each buffer to reduce observation noise.

\noindent
The “koopman\_MPC” node computes Koopman-based predictions and solves the MPC QP. Observable matrices are constructed via the lifting function $\psi(\cdot)$

\begin{equation}
    {\mathcal{O}^{[i]}}(k) = [\psi(\mathcal{H}^{[i], 1}(k)), \dots, \psi(\mathcal{H}^{[i], O^{[i]}(k)}(k))],
\end{equation}

\noindent
$O^{[i]}(k)$ is kept between 5 and 25 to ensure stable prediction. The Koopman operator is recomputed every $T_\kappa=1 \mathrm{s}$ according to:

\begin{align}
    \widetilde{\mathcal{K}}^{[i]}(k) &= \mathcal{O}^{[i], 2:O^{[i]}(k)}(k) \times \mathcal{O}^{[i], 1:O^{[i]}(k)-1}(k)^\dagger, \\
    \mathcal{K}^{[i]}(k) &= \left(\widetilde{\mathcal{K}}^{[i]}(k)\right)^{\frac{{T_s}}{T_{\kappa}}}.
\end{align}



Here, $A^\dagger$ stands for the pseudo-inverse of matrix $A$. 

\noindent
Future positions are then recursively estimated as:

\begin{align}
    \mathcal{H}_\mu^{[i]}(k) &= \psi^{-1}(\mathcal{K}^{[i]}(k) \times \psi(\mathcal{H}_{\mu-1}^{[i]}(k))), \\
    p_\mu^{[i]}(k) &= \mathcal{H}_\mu^{[i], {1:3}}(k), ~~ \forall \mu \in \mathbb{N}_{[1, H - 1]}.
\end{align}


\noindent
The MPC uses a dodecahedral polytope for obstacle representation. The velocity controller uses $\bm{K_{vel}}=1.8\mathbf{I}^{3 \times 3}$ (matching PX4 defaults), and CVXPY is used to solve the QP in under $0.1 \mathrm{s}$ on average.
For prediction validation, we test three lifting functions $\psi_{\text{p}}, \psi_{\text{pv}}, \psi_{\text{pva}}$ which approximate the instantaneous position; position and velocity; and position, velocity and acceleration of the UAV. Two obstacle trajectories were tried: circular and infinity-shaped. Results, shown in Figure \ref{FIG_pred_results_1}, confirm that velocity and acceleration information enhance prediction accuracy for curved paths.

\noindent
Prediction accuracy versus history size and prediction step is shown in Figure \ref{FIG_pred_results_2}. A history of 10 vectors provides a good trade-off between accuracy and efficiency.
For full-system evaluation, a mission with 3 moving obstacles is simulated. Figure \ref{FIG_3D_avoidance} shows successful navigation toward the goal while avoiding collisions.
Figure \ref{FIG_distance_to_obstacle} confirms that the safety distance is respected throughout the mission. Further illustration in Figure \ref{FIG_2D_avoidance} shows effective avoidance with prediction alignment. Table \ref{tab:error_comparison} presents prediction errors comparing our Koopman method with Gaussian Processes (GP) as used in \cite{olcay2024dynamic}, showing superior performance in all metrics. Table \ref{tab:time_diff} highlights the dramatic computational speedup (98\%) offered by the Koopman-based method over GP, which is vital for real-time systems.

\section{Conclusion}
\label{Sec. 6}

This work has presented a Koopman-based MPC framework for real-time UAV navigation, specifically addressing the challenge of obstacle avoidance in dynamic environments. By leveraging the Koopman operator to approximate nonlinear dynamics through a linear representation and integrating this with real-time LiDAR data, we developed a control strategy that is both computationally tractable and effective in practice. The proposed approach has been validated within the ROS2-Gazebo simulation environment, demonstrating reliable performance in scenarios involving moving obstacles, and confirming its applicability under realistic sensing and actuation conditions.
A compelling direction for future research is the incorporation of stochastic modeling into the Koopman-based framework. Since the motion of dynamic obstacles is inherently uncertain due to noisy sensor measurements, limited visibility, or unmodeled behavior, modeling the Koopman operator as a stochastic system would offer more robust and realistic prediction capabilities. This extension could further enhance the safety and adaptability of UAV operations in uncertain and rapidly changing environments.

\bibliographystyle{IEEEtran}
\bibliography{Refrences}

\end{document}